\documentclass[usenatbib,fleqn]{mnras}

\usepackage{newtxtext,newtxmath}
\usepackage[T1]{fontenc}
\usepackage{ae,aecompl}

\usepackage{amsmath}	% Advanced maths commands
\usepackage{amssymb}	% Extra maths symbols

\usepackage{ctable}
\usepackage{url}
\usepackage{xspace}
\usepackage[normalem]{ulem}
\def\msun{{\rm\,M_\odot}}

\newcommand{\be}{\begin{equation}}
\newcommand{\ee}{\end{equation}}
\newcommand{\der}[2]{\ensuremath{\frac{{\rm d} #1}{{\rm d} #2}}}
\def\h2{${\rm\,H_2}$}

\title[PBHs as magnetic field seeds]{Primordial black holes as seeds of magnetic fields in the universe}
\author[M. Safarzadeh]{Mohammadtaher Safarzadeh\thanks{E-mail: mts@asu.edu}\\
	School of Earth and Space Exploration, Arizona State University, Tempe, AZ 85287-1404, USA;
}

\begin{document}

\pubyear{2017}

\maketitle

\label{firstpage}

\begin{abstract} 

Although it is assumed that magnetic fields in accretion disks are dragged from the interstellar medium, the idea
is likely not applicable to primordial black holes (PBHs) formed in the early universe.
Here we show that magnetic fields can be generated in initially unmagnetized accretion disks around PBHs through the Biermann battery mechanism, and therefore provide the small scale seeds of magnetic field in the universe.
The radial temperature and vertical density profiles of these disks provide the necessary conditions for the battery to operate naturally. 
The generated seed fields have a toroidal structure with opposite sign in the upper and lower half of the disk.
In the case of a thin accretion disk around a rotating PBH, the field generation rate increases with increasing PBH spin. 
At a fixed $r/r_{\rm isco}$, where $r$ is the radial distance from the PBH
and $r_{\rm isco}$ is the radius of the innermost stable circular orbit, 
the battery scales as $M^{-9/4}$, where $M$ is the PBH's mass. 
The very weak dependency of the battery on accretion rate, makes this mechanism a viable candidate to provide seed fields in an initially unmagnetized accretion disk, following which the magnetorotational instability could take over. 

\end{abstract}
 

\section{Introduction}

The origin of magnetic fields observed in galaxies remains elusive and various mechanisms such as density perturbations before recombination era \citep{Ichiki:2006cd} or linear over-densities \citep{Naoz:2013er} have been proposed to generate large scale seeds in the universe. On small scales, seeds are proposed to be formed 
through the Biermann battery mechanism \citep{Biermann} in stars and supernova explosions \citep{Widrow:2002ke,Hanayama:2005fb}.

Here we turn our attention to the primordial black holes (PBHs) that form due to density fluctuations in the early universe \citep{CH1974,Carr:1975gg}.
Interest in PBHs has recently rekindled \citep{Bird2016} following the detection
of two 30 $\msun$ black holes \citep{Abbott:2016ki} by LIGO. Through various approaches such as micro lensing \citep{2001ApJ...550L.169A}, CMB anisotropy \citep{AliHaimoud:2017hy}, wide binaries \citep{Quinn:2009kc,MonroyRodriguez:2014iv}, and ultra faint dwarfs \citep{Brandt:2016ga}, PBHs in the mass range of 10-100 $\msun$ are considered to potentially play an important role in cosmology, although they cannot account for the major part of dark matter density.

While it is predominantly assumed that PBHs grow by accreting primordial gas through Bondi spherical accretion \citep{Ricotti:2008kq,AliHaimoud:2017hy}, or accretion of dark matter \citep{Mack:2007hy}, 
here we show the possibility that these black holes could grow via an accretion disk and how this will generate the small scale seeds of magnetic fields in the early universe.

Accretion disks are associated with a wide range of astrophysical phenomena from disks around protostars to BHs. They are assumed to be magnetized
and the accretion to occur via the magnetorotational instability (MRI, \citealt{BlbusHawley}). 
Generally, large-scale magnetic fields are assumed to be dragged with the accretion flow, and amplified by dynamo process during the accretion into the disk \citep[e.g.,][]{Bisnovatyi+76,Reyes+96,Calvet98,Reyes+99,Blackman+01}.
For stellar mass BHs, the magnetic field in the accretion disk is typically associated with the magnetic field of the progenitor star \citep[e.g.,][]{Bisnovatyi+74,Bisnovatyi+76}, 
while for supermassive BHs the field is believed to be accreted from the surrounding gas \citep[e.g.,][]{mad,Eatough}.
Magnetic fields around the first generation of protostars may be the result of large scale cascades, where weak magnetic fields are amplified via a turbulent dynamo during the gravitational collapse \citep[e.g.,][]{Schleicher+10,Schober+12}.

While it remains unknown how effectively those sources can provide 
magnetic fields around PBHs during different cosmological epochs,
here we focus on generating seed magnetic fields in initially non-magnetized accretion disks around PBHs.
We do not specify the cause of accretion in the system; we merely assume that it takes place despite the fact that the disk is {\it not} initially magnetized (i.e., MRI does not yet operate). 
Such disks have been studied in the protoplanetary (PP) disk context in detail \citep{CG1997,CG1999,Dullemond07} and it was shown that a pure hydrodynamic setup can lead to significant accretion rates in astrophysical accretion disks \citep{Paoletti+12}.

\citet{Biermann} showed that if a plasma has rotational motion, currents must exist, which lead to the generation of magnetic fields. 
This process has a wide range of applications from the generation of magnetic fields in stars \citep[e.g.,][]{Biermann,doi} to galactic scale magnetic fields \citep[e.g.,][]{Mestel+62,Widrow:2002ke,Subramanian+94,Subramanian10,Naoz:2013er}.
For example, \citet{Shiromoto+14} studied the generation of magnetic fields 
by the Biermann battery mechanism in a first generation proto-circumstellar disk based on 2D radiation hydrodynamics simulations.

In order to generate seed magnetic fields via the Biermann battery mechanism, the gradient of electron density and gradient of electron pressure must have a non-vanishing cross product.
This situation naturally arises in accretion disks, where the pressure gradient is mostly in the vertical direction, while the temperature has 
a radial gradient. The seed magnetic field via the battery will grow linearly with time. Subsequently, this could trigger a strong magnetohydrodynamic (MHD) 
instability through the MRI, which would cause the field to grow exponentially with time to reach its equipartition value in the disk \citep[e.g.,][]{Bret09}.
As we will show below, generation of the seed magnetic fields through the Biermann battery depends only \emph{very weakly} on the accretion rate and
therefore a priori we do not require MRI to operate for the battery to work. 

The structure of this work is as follows:
In \S2 we introduce the Biermann battery. 
In \S3 we apply the battery to a thin accretion disk around a Kerr BH. In \S4 we present our results, and in \S5 we present our conclusions.

\section{Biermann Battery}

%We begin by describing the general Biermann battery mechanism. 
The generation of a magnetic field via the Biermann battery is due to currents
driven in a plasma whenever there is a non-vanishing cross product of electron number density gradient, $\nabla {n_e}$, and electron pressure gradient, $\nabla P_e$,
\be
\label{eq:Bier}
\frac{\partial {\bf B}}{\partial t}= \nabla \times \left( {\bf u} \times {\bf B}\right) -  c \frac{\nabla n_e \times \nabla P_e}{e n_e^2} \ .
\ee
Assuming an ideal gas equation of state, $P_e=n_e k_B T_e$, and assuming that the temperature of the electrons is equal to that of the gas, we can re-write ${\nabla n_e \times \nabla P_e}$ as
\be
\nabla n_e \times \nabla P_e = \nabla n_e \times  (n_e\nabla T_e+ T_e \nabla n_e) k_B \ ,
\ee
where $k_B$ is the Boltzmann constant. Given $n_e=\chi_e \rho$, where $\chi_e$ is the ionization fraction, we find
\be
\nabla n_e \times \nabla P_e = n_e k_B \nabla n_e \times \nabla T_e = n_e k_B \chi_e \nabla \rho \times \nabla T_e \ .
\ee

We assume the disk is initially not magnetized. We have assumed $\nabla \chi_e=0$ considering the ionization fraction
remains effectively unchanged at the relevant temperatures of accretion disks.
Therefore, we estimate the rate of growth of the magnetic field to be
\be
\label{e:4}
\frac{\partial {\bf B}}{\partial t}\approx- \frac{c k_B }{e } \frac{\nabla \rho \times \nabla T_e }{\rho}\;.
\ee
Note that the ionization fraction, $\chi_e$, cancels out, and the result depends only on the density and temperature of the system. 
Since $\nabla \rho$ and $\nabla T_e$ are both poloidal (by axisymmetry), the generated field is toroidal, i.e.,
\be\label{eq:dBdt}
\frac{\partial {\bf B}}{\partial t} =-\frac{c k_B }{e \rho} \left(\frac{\partial T_e}{\partial r} \frac{\partial \rho}{\partial z}- \frac{\partial \rho}{\partial r} \frac{\partial T_e}{\partial z}\right){\hat \phi}\;.
\ee

The Biermann battery will work so long as the density and temperature
gradients are not parallel to each other. This condition will almost
always be satisfied in an accretion disk. 
%explain why the vertical structure is ignored. 

%%%%%%%%%%%%%%%%%%%%%%% Magnetic seeds in disks around Kerr black holes          %%%%%%%%%%%%%%%%%%%%%%%%%%%%%%%%%%%%%
\section{Magnetic seeds in disks around Kerr black holes}
In this section we explore the application of the Biermann battery to Kerr black holes which is relevant for PBHs since they tend to form with high spin parameters \citep{Harada:2017cq}.
The classical disk model of \citet{ss} was generalized to a relativistic disk by \citet{nt}, in which the condition of vertical hydrostatic equilibrium
takes the form
\be
 \der Pz=-\rho \frac{GM}{r^3} z \frac{\cal C}{\cal B} \ ,
\ee
where $M$ is the mass of the central BH. ${\cal B}$, and ${\cal C}$
are relativistic correction factors. This leads to the following density profile
\be
\rho=\frac{\Sigma}{\sqrt{2 \pi}H} \exp(\frac{-z^2}{2 H^2}\frac{\cal C}{\cal B}) \sqrt{\frac{\cal C}{\cal B}} \ ,
\ee

where $H$ is the disk scale height defined as $H=C_s/\Omega$. $C_s=\sqrt{k_B T_{\rm eff}(r)/(\mu m_p)}$ is the speed of sound, and $\Omega=\sqrt{G M/r^3}$ is the Keplerian angular velocity. 
The surface density ($\Sigma$) is given by
\be
\Sigma=\frac{\dot M}{3 \pi \nu} \left(1-\sqrt{\frac{r_{\rm isco}}{r}}\right) \ ,
\ee

where $\nu$ is the viscosity coefficient, $\dot M$ is the accretion rate, and $r_{\rm isco}$ is the radius of the innermost stable circular orbit. 
Both $H$ and $\nu$ are functions of $r$ and $a$, where $a$ is the specific angular momentum of the BH. 

The effective temperature profile in the disk from the balance of heating and cooling is given by
\be
T^4_{\rm eff}(r) =\frac{3GM{\dot M}}{8\pi \sigma r^3} \left(1-\sqrt{\frac{r_{\rm isco}}{r}}\right) \frac{{\cal D}}{{\cal B}} \ ,
\ee
where $\sigma$ is the Stefan-Boltzmann constant, and ${\cal D}$ is a relativistic correction factor. 
We compute the battery at $z=H$ for all the calculations as a reference. 
In all the analysis, we set $T_e = T_{\rm eff}$
\footnote{The disk mid-plane temperature is often just a factor of a few larger than $T_{\rm eff}$. We feel therefore that this approximation is reasonable.}. 
The various relativistic factors are given by \citep{nt,pagethorne,doerrer}:
%{\color{green} (is ${\cal A}$ used anywhere? If not, eliminate the equation and make sure to check that all the equation numbers are okay after the change)}
\begin{eqnarray}
%{\cal A}&=& 1-\frac{2GM}{c^2 r} +\frac{a^2}{c^2 r^2}                \nonumber \\
{\cal B}&=& 1-\frac{3GM}{c^2 r} +\frac{2a\sqrt{GM}}{c^2 r^{3/2}},    \nonumber \\
{\cal C}&=& 1-\frac{4a\sqrt{GM}}{c^2 r^{3/2}} +\frac{3a^2}{c^2 r^2}, \nonumber \\
{\cal D}&=& \frac{1}{2\sqrt{r}} \int_{r_{\rm isco}}^r
            \frac{x^2 c^2 - 6xGM + 8a\sqrt{xGM} - 3a^2}
                 {\sqrt{x}\left( x^2 c^2 - 3xGM + 2a\sqrt{xGM} \right)}\ dx,  \ 
\end{eqnarray}
and $r_{\rm isco}$ is obtained as the root of the following equation for the case of a co-rotating disk:
\be
1-6\frac{MGc}{r} + 8 \frac{a}{c} \sqrt{\frac{MGc}{r^3}}-3 \frac{a}{c r}=0 \ .
\ee

The vertical temperature profile of the disk depends on the assumed opacity of the disk and for an optically thick, geometrically thin disk we have:
\be
\frac{8\sigma T_c^4}{\kappa_R \Sigma}=\frac{-4\sigma}{3 \kappa_R \rho}\frac{\partial T_e^4}{\partial z}
\ee
where $\kappa_R$ is the Rosseland mean opacity and $T_c$ is the mid-plane temperature of the disk, related to the effective temperature by:
\be
T_c^4=\frac{3\tau}{8}T_{\rm eff}^4
\ee
Computing the vertical gradient of temperature at scale height of the disk, for $\tau=10^{3}-10^{4}$, we get:
\be\label{eq:dTdz}
\frac{\partial T_e}{\partial z} \approx -\tau \frac{T_{\rm eff}}{H}
\ee

Due to large uncertainties in the vertical temperature profile of the disk, we
estimate the order of magnitude of the Biermann battery by only crossing
the vertical gradient of density with the radial gradient of
temperature for Kerr black holes.
Therefore, we compute ${\partial {\bf B}}/{\partial t}$ as : 
\be\label{eq:Bier2}
\frac{\partial {\bf B}}{\partial t}\approx-\frac{c k_B }{e \rho} \left(\frac{\partial T_e}{\partial r}\frac{\partial \rho}{\partial z}\right){\hat \phi} \ .
\ee

%%%%%%%%%%%%%%%%%%%%%%% Results    %%%%%%%%%%%%%%%%%%%%%%%%%%%%%%%%%%%%%

\section{Results}

Here we explore thin accretion disks around PBHs with a constant vertical temperature profile (thus, Equation (\ref{eq:Bier2}) is applicable).

\begin{figure}
\resizebox{3.3in}{!}{\includegraphics{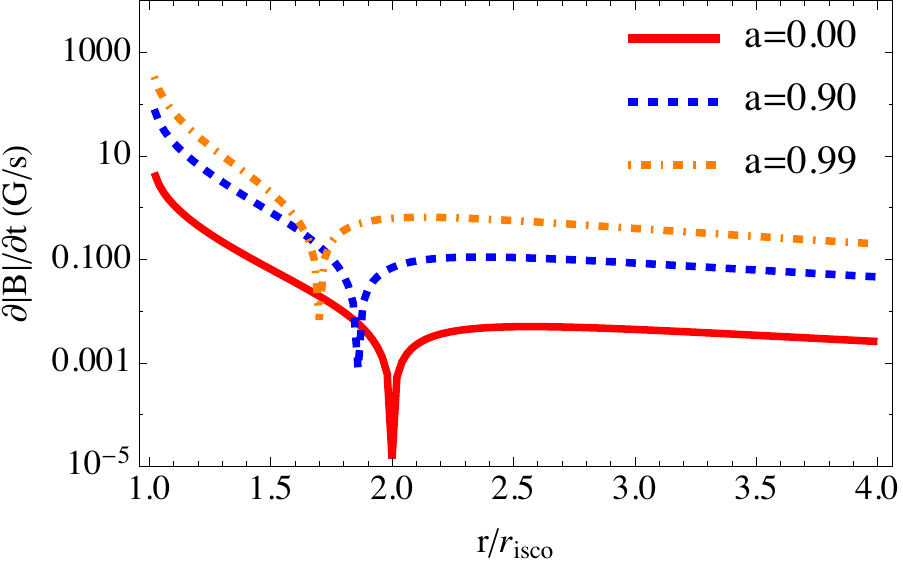}}
\caption{The magnetic field generation rate (absolute magnitude) as a function of $r/r_{\rm isco}$ for a $10\,M_{\odot}$ PBH with
accretion rate of $\rm \dot M= 10^{-10}M_{\odot}/year$ for different spin parameters. The geometry of the generated fields is toroidal. We have set $z= H=C_s/\Omega$ in this calculation for reference. 
The battery is about 2 orders of magnitude stronger for $a=0.99$ (rapidly spinning Kerr BH) compared to $a=0$ (Schwarzschild BH). The notches in the
curves indicate radii $r_{\rm turn}$ at which the generated seed fields change sign. We observe that $r_{\rm turn}=2\times r_{\rm isco}$ for a Schwarzschild BH and decreases with increasing spin. 
At each radius, the field changes sign across the equatorial plane.}\label{fig:test_a}
\end{figure}

Figure \ref{fig:test_a} shows the magnetic field growth rate for a $10~M_{\odot}$ PBH with different spin parameters ($a$) as a function of $r/r_{\rm isco}$. 
We compute the battery term at $z=H=C_s/\Omega$ for reference. We see that the battery is stronger for PBHs with larger spin parameter at a fixed $r/r_{\rm isco}$.
The seed fields generated by the Biermann battery have opposite sign in the upper and lower half of the disk. In addition, the fields
change sign at a radius $r_{\rm turn}$ (indicated by the sharp dips in the curves) because $T_e$ has a maximum at this radius. 
$r_{\rm turn}=2\times r_{\rm isco}$ for a Schwarzschild BH and at smaller radii for Kerr BHs.
Such a maximum is predicted by the thin disk model \citep{ss,nt}, though the existence of the maximum in
a real disk is debated (see \citealt{2008ApJ...687L..25S,2011MNRAS.414.1183K} for a discussion of the zero-torque condition at the ISCO and the effect it has on the viscous heating profile in a thin accretion disk).
The battery scales with radial distance from the central BH as $\propto r^{\gamma}$ where $\gamma\approx-2.5$ at $r \gtrapprox 2 \times r_{\rm isco}$.
We estimate the generated magnetic field as $B=\frac{\partial {\bf B}}{\partial t}/\Omega$, where we take $t_{\rm dyn}\propto 1/\Omega$ as the dynamical timescale for a PBH.
The generated $B$ for a $10 \msun$ PBH is $\sim 1.5\times 10^{-5} G$ at $r\approx4\times r_{\rm isco}$ and scaling approximately as $\propto r^{-1}$.
Such large amplitude of the generated B field could be compared to what is predicted around the first generation stars by radiation pressure on the order of $10^{-19} G$ \citep{Ando2010}.

\begin{figure}
\resizebox{3.2in}{!}{\includegraphics{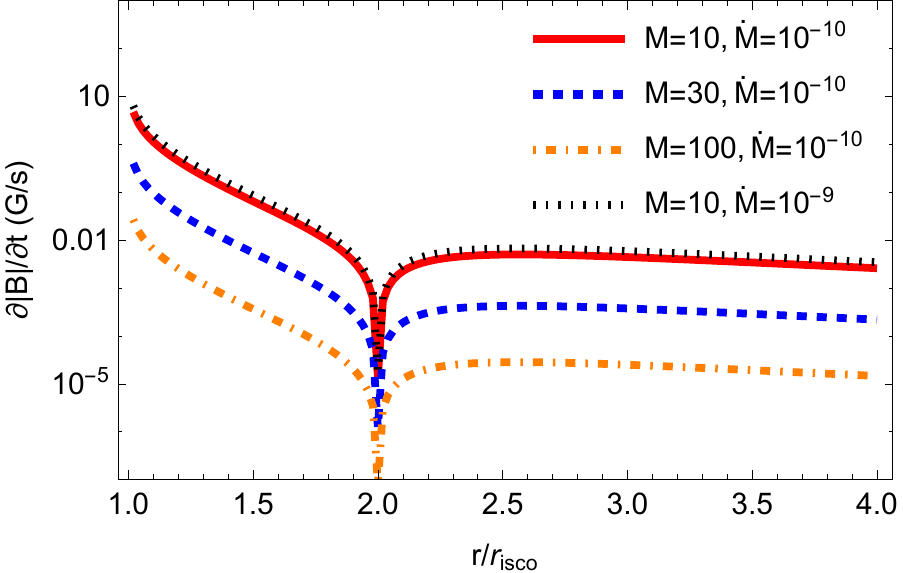}}
\caption{The absolute magnitude of magnetic field generation rate as a function of $r/r_{\rm isco}$ for a Schwarzschild PBH with masses and accretion rates in $\msun$ 
and M$_{\odot}/$year units respectively. The red solid line corresponds to $M=10~\msun$ and $\rm \dot M= 10^{-10}~M_{\odot}/year$.
The blue dashed (orange dot-dashed) line corresponds to  $30~(100)~\msun$ PBH with the same accretion rate respectively. 
The dotted black line corresponds to $M=10~\msun$ with an order of magnitude larger accretion rate. At a fixed $r/r_{\rm isco}$ the battery is proportional to $M^{-9/4}$ and is weakly dependent on accretion rate ($\propto \rm \dot M^{1/8}$). 
The radial turning point ($r_{\rm turn}$) is independent of both mass and accretion rate. We assume $z= H$ in this calculation.}\label{fig:dBMdot}\vspace{0.2cm}
\end{figure}

Figure \ref{fig:dBMdot} shows the growth rate of magnetic field for PBHs with different masses and accretion rates. 
The battery is proportional to $M^{-9/4}$ at a fixed $r/r_{\rm isco}$, but is only weakly dependent on the accretion rate ($\propto \dot M^{1/8}$ at $z=H$). 
For example, for a $10~\msun$ PBH, at $r\approx4\times r_{\rm isco}$, the battery rate is around $2\times10^{-3} G/s$ with very weak dependence on the accretion rate.
For an order of magnitude more massive black hole the battery rate is $\approx2\times10^{-5} G/s$ which is about two orders of magnitude smaller.

\begin{figure}
\resizebox{3.3in}{!}{\includegraphics{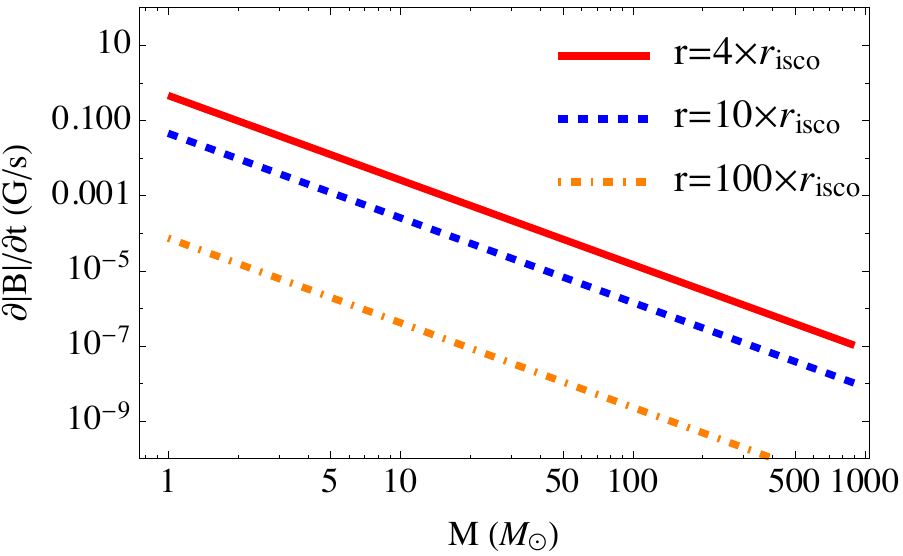}}
\caption{The absolute magnitude of magnetic field generation rate as a function of the Schwarzschild PBH mass at three different radial distances from the BH. 
The battery scales as $M^{-9/4}$ independent of the radial distance from the BH. We assume $z= H$ in this calculation.}\label{fig:M_study}
\end{figure}

We show the dependence of the battery on the PBH mass in Figure \ref{fig:M_study}. We find ${\partial {\bf B}}/{\partial t} \propto M^{-9/4}$ at all radii from the PBH.

%%%%%%%%%%%%%%%%%%%%%%%%%%%%       Discussion           %%%%%%%%%%%%%%%%%%%%%%%%%%%%%%%%%%%%%

\section{Summary and Discussion} 
We showed that toroidal magnetic fields could naturally be generated through the Biermann battery process starting initially from zero magnetic field in accretion disks around PBHs.
The vertical density profile of an accretion disk and the radial temperature profile which arises due to the balance between heating and cooling, 
lead to generation of toroidal seed magnetic field through the battery. 

We adopted a disk temperature and density profile, which includes relativistic corrections \citep[following][]{nt,pagethorne,doerrer}. The generated seed fields change their sign in the azimuthal direction at $r_{\rm turn}=2\times r_{\rm isco}$ for a Schwarzschild BH. Larger BH spins yield smaller values of $r_{\rm turn}$ and larger rate of seed generation (see Figure \ref{fig:test_a}). We note that the vanishing of $T_e$ at $r=r_{\rm isco}$, \citep[as predicted by][]{ss} is controversial because it comes from the zero-torque boundary condition at the ISCO and from ignoring advection. Thus, the behavior of sign flip may not be realistic. 
At a fixed $r/r_{\rm isco}$ the battery scales as $M^{-9/4}$. The relativistic disk profile leads to a weak dependency on the BH accretion rate $\propto \dot M^{1/8}$ (see Figure \ref{fig:dBMdot}).   

While the presence of magnetic fields around PBHs from the cascade of large scale seeds that are potentially generated during the inflationary period remains unknown, our proposed mechanism introduces an alternative source
 as early as the formation of the first PBHs. Moreover, the generated seeds alter the growth mode of PBHs beyond Bondi accretion with implications to be studied further in the future.

\section{Acknowledgements}
MS is thankful to the referee for their constructive comments. MS is also thankful to Smadar Naoz and Lorenzo Sironi for useful discussions.
MS is thankful to Arif Babul and Evan Scannapieco for their comments on an earlier draft of this work.
MS is supported by the National Science Foundation under grant AST14-07835 and by NASA under theory grant NNX15AK82G. 

\bibliographystyle{mnras}
%\bibliography{the_entire_lib}

\bsp
\label{lastpage}
\end{document}